\def\simgt{\lower.5ex\hbox{$\; \buildrel > \over \sim \;$}}
\def\simlt{\lower.5ex\hbox{$\; \buildrel < \over \sim \;$}}

\newcommand\teff{T$_{\rm eff}$}

\newcommand{\msun}{\ensuremath{\, {M}_\odot}}
\newcommand{\Msun}{\ensuremath{\, {M}_\odot}}
\newcommand{\ocen}{$\omega$~Cen}

\newcommand{\Mrg}{M$_{\rm RGB}$}

\newcommand{\Mturb}{M$_{\rm turb}$}

\newcommand{\ysurf}{Y$_{\rm surf}$}
\newcommand{\yin}{Y$_{\rm in}$}

\newcommand{\col}{m$_{\rm F336W}$--m$_{\rm F814W}$}

\documentclass[useAMS,usenatbib]{mnras}
\usepackage{graphicx}
\usepackage{epstopdf}
\usepackage{txfonts}

\usepackage[T1]{fontenc}
\usepackage{ae,aecompl}

\usepackage[usenames,dvipsnames]{color}

\title[HB morphology and diffusion in NGC 6388 and NGC 6441]{The Hubble Space Telescope UV Legacy Survey of Galactic Globular Clusters  -- XI.  The 
horizontal branch in NGC\,6388 and NGC\,6441\thanks{Based on observations with the NASA/ESA Hubble Space Telescope, obtained at the Space Telescope Science Institute, which is operated by AURA, Inc., under NASA contract NAS 5-26555.}}
\author[Tailo et al.]{M. Tailo$^{1}$\thanks{E-mail:
mrctailo@gmail.com (MT); franca.dantona@gmail.com (FDA)},
F. D'Antona$^{1}$, A. P. Milone$^{2}$, A. Bellini$^{3}$, P. Ventura$^{1}$, M. Di Criscienzo$^{1}$, 
\newauthor    S. Cassisi$^{4}$,  G. Piotto$^{5,6}$, M. Salaris$^{7}$, T.M. Brown$^{3}$, E. Vesperini$^{8}$,  L. R. Bedin$^{6}$, \newauthor A.F. Marino$^{2}$, D. Nardiello$^{5}$,  and J. Anderson$^{3}$
\\ \\
$^{1}$INAF- Osservatorio Astronomico di Roma, via di Frascati 33, I-00040 Monteporzio (Italy) \\ 
$^{2}$  Research School of Astronomy \& Astrophysics, Australian National University, Canberra ACT 2611, Australia\\
$^{3}$  Space Telescope Science Institute, 3700 San Martin Drive, Baltimore, Maryland 21218, USA\\
$^{4}$ Osservatorio Astronomico di Teramo, Via Mentore Maggini s.n.c., I-64100 Teramo, Italy;\\
$^{5}$ Dipartimento di Fisica e Astronomia ``Galileo Galilei",  Universit\`a di Padova, Vicolo dell'Osservatorio 3, I-35122 Padova, Italy \\
$^{6}$ INAF-Osservatorio Astronomico di Padova, Vicolo dell'Osservatorio 5, I-35122 Padova, Italy;\\
$^{7}$  Astrophysics Research Institute, Liverpool John Moores University, Liverpool Science Park, IC2 Building, Liverpool L3 5RF, UK\\
$^{8}$ Department of Astronomy, Indiana University, Swain West, 727 E. 3rd Street, IN 47405 Bloomington (USA)\\
}
\begin{document}

\pagerange{\pageref{firstpage}--\pageref{lastpage}} \pubyear{2002}

\maketitle

\label{firstpage}

\begin{abstract}
The \textit{Hubble Space Telescope} UV Legacy survey of Galactic Globular Clusters (GC) is characterising many different aspects of their multiple stellar populations.  The ``Grundahl--jump" (G--jump) is a discontinuity in ultraviolet brightness of blue  horizontal branch (HB) stars, signalling the onset of radiative metal levitation. The HB Legacy data confirmed that the G--jump is located at the same \teff\ ($\simeq$11,500\,K) in nearly all clusters. The only exceptions are the metal--rich clusters NGC\,6388 and NGC\,6441, where the G--jump occurs at  \teff$\simeq$13-14,000K. 
We compute synthetic HB models based on new evolutionary tracks including the effect of helium diffusion, and approximately accounting for the effect of metal levitation in a stable atmosphere. Our models show that  the G--jump location depends on the interplay between the timescale of diffusion and the timescale of the evolution in the \teff\ range 11,500\,K$\lessapprox$\teff$\lessapprox$14,000\,K. The G--jump becomes hotter than 11,500\,K only for stars that have, in this \teff\ range, a helium mass fraction Y$\simgt$0.35. Similarly high Y values are also consistent with the modelling of the HB  in NGC\,6388 and NGC\,6441. In these clusters we predict that a significant fraction of HB stars show helium in their spectra above 11,500\,K, and full helium settling should only be found beyond the hotter G--jump.
\end{abstract}


\section{Introduction}
Two problems cohexist in determining the stellar distribution along the horizontal branch (HB) of Globular Clusters (GCs): the presence of {\it discrete} multiple populations in practically all clusters  \citep[see, for a recent update, Paper I and Paper IX,][]{piotto2015, milone2016} and the presence of phenomena ascribed to well identified variations in the atmospheric composition of the stars. 
In particular, the `Grundahl jump' (G--jump) in the Str\"omgren $u$\ color \citep{grundahl1998} has been shown to be connected to the sudden onset (at  \teff$\sim 11,500$\,K) of enhanced abundances of elements heavier than carbon and nitrogen in the atmospheres of blue HB stars hotter than this \teff. 
When the envelope is free from any form of turbulence, some elements suffer radiative levitation, due to radiation pressure on resonant lines, and the spectrum changes due to the higher concentration of metals. The result is a sudden `jump' in the magnitudes and colors most affected by metal lines \citep{grundahl1999}. 
A concomitant decrease in the helium abundance is observed in the atmospheres of stars hotter than $\simeq$11,500\,K, while cooler stars have normal helium. Typically, the helium to hydrogen ratio becomes a factor 10--100 smaller than the solar value 
\citep{behr2003, monibidin2009, monibidin2012}.  Blue HB stars also show a drop in rotation rates close to the G--jump \citep[e.g.][]{behr2000, recioblanco2004}.

The reasons for the presence of another discontinuity at \teff$\sim$20,000\,K is less clear \citep{momany2004}, as discussed in paper VII by \citet{brown2016}, while the most extreme gap, at \teff$\simeq$35,000\,K is interpreted as the \teff\ discontinuity between the end of standard HB and the ``late--flashers" stars, in which helium is the dominant atmospheric constituent \citep{sweigart1997, moehler1999, brown2001}.  

The main \teff\ location of the HB in different clusters depends on age and metallicity,  while its color extension has been satisfactorily explained as the primary effect of the spread in the initial helium content (\yin) of the cluster populations \citep{dantona2002}\footnote{The reader must keep in mind that the helium content has two different aspects and meanings in this work: 1) {\it the ``initial" helium mass fraction Y, or \yin} in the evolving stars. This value may be different for the different ``multiple populations". It affects the stellar evolution (lifetimes, luminosities, \teff). In HB, it affects  the morphology of the evolutionary track).  2) {\it The surface helium mass fraction \ysurf}. Starting from a specific \yin\ value of the evolving star, we consider how the helium abundance is modified by  diffusion and mixing processes in the stellar envelope, and in particular at the surface.}.
The larger is the \yin, the hotter is the HB location of core--helium burning stars, even if the mass loss during the previous red giant branch phase is the same.
A simple argument in favour of the role of helium is that the clusters showing the most relevant abundance anomalies also have the HBs most extended in color  \citep{gratton2010, milone2014}, and self--enrichment models predict a direct correlation between light elements and helium content anomalies.  One of the main chemical signatures of second generation, the sodium abundance, on average is larger at hotter HB locations \citep[e.g.][]{marino2011m4, gratton2011hb2808, milone47tuc2012, gratton2014} in agreement with the expectation that higher Y HB stars have larger \teff. 
The presence of very large helium contents (Y$\gtrsim$0.35) in the stars populating the `blue hook', a group especially relevant in a few massive clusters\footnote{See  \cite{brown2016}, paper VII, for an update of the blue hook presence in clusters from the UV Legacy Survey observations.}  is more clearly confirmed by the presence of a similar high helium {\it main sequence} population in \ocen\ \citep{bedin2004,norris2004} and NGC\,2808 \citep{dantona2005, piotto2007}. 
\\
Anyway, HB morphologies can not be characterized solely in terms of helium abundance. Two other parameters play a role in the location of stars along the HB: 1) the possible increase in the total CNO content in some populations in a few clusters \citep{marino2013m22hb, dantona2016}, which shifts the tracks to {\it cooler} locations \citep[e.g.][]{salaris2008}, and  2) mass loss in the red giant branch, which may be larger in second generation stars \citep[see the discussion in][]{dantona2013}, shifting the tracks to {\it hotter} locations. 
The data acquired by the Hubble Space Telescope Treasury survey (paper I), are providing strong evidence that, in most clusters, multiple stellar populations form discrete groups differing in the mean average chemical composition of their member stars (paper IX). In this case, it is probable that also the helium content is discontinuous passing from one group to the other, although the differences between contiguous groups may be very small in Y.

It is not clear whether {\it helium discontinuities} are necessary to explain any of the discontinuities in the HB. 
Paper VII reconsidered the morphology of HB using the photometric catalogues from the UV Treasury survey (paper I),  and established beyond any doubts that the three important discontinuities cited above are located at the same \teff\ in nearly all clusters, and therefore can not be related to differences in the initial Y among specific groups of progenitor stars. Discontinuities at other \teff\ locations should be examined one by one to assess the possible role of helium differences, as it would be interesting to understand whether the morphology of HBs provides clear evidence for helium differences due to the presence of multiple populations. In some specific cases, this problem has been already addressed. \cite{dc2004} established that a helium discontinuity was responsible for the scarcity of stars in the RR~Lyr region of NGC\,2808, an hypothesis that has been subsequently confirmed by the discovery of its triple main sequence.  \cite{dicrisci2011b} and \cite{dicrisci2015} established that the stars in NGC\,2419 are mainly distributed into a group with standard Y and another one with very high Y.

\cite{brown2016} have identified  the G--jump at \teff$\sim$13,000--14,000K in the metal rich peculiar clusters NGC\,6441 and NGC\,6388. They attribute the presence of a discontinuity at such a hotter \teff\ to the possibility that this \teff\ region is populated by extremely helium rich stars (\yin$\simgt$0.35). The high--\yin\ stars are confined to the hotter end of the HB and/or to the blue hook stars in other clusters where they are found. Their location at such, much lower, \teff\ in NGC\,6441 and NGC\,6388 is in agreement with the identification of the presence of high--\yin\ stars also in the thick red clump  and in the RR-Lyrae strip, where they cause of long, anomalous periods\citep{caloidantona2007}. 

In this work we compare the observed HB of NGC6388 and NGC6441 with new HB models that include diffusion. To do this, we will use a library of new HB evolutionary tracks  \citep[see ][for the application of these models to the study of \ocen]{tailo2015,tailo2016}.
We show here that models including helium diffusion and helium contents Y $\sim$\ 0.35 -- 0.38 are necessary to explain the hotter Grundahl's jump in these clusters. We predict that a relevant fraction of the stars populating the region $11,500 \simlt$\teff$\simlt 13,000$K must show non negligible \ysurf\ in their atmosphere, and that full helium settling occurs only in stars of larger \teff. 
 In further coming papers we will use the same sets of  models to infer the helium distribution and the mass loss of HB stars in the GCs studied in the Treasury UV survey (Tailo et al. 2017a,b, in preparation). 

The outline of the paper is the following. In \S~\ref{sec2} we discuss the observations of the two clusters, and their relevance in the study of multiple populations in GCs, in particular concerning helium variations. 
In \S~\ref{sec3} we explain the input physics of the stellar models employed in this work, and, in particular, the formulation of the helium diffusion parametrization. We compare the high--Y model excursions through the HR diagram in models with and without helium settling. In  \S~\ref{sec4} we show the results of synthetic HB models and their comparison with the Treasury data for the two clusters. 
In \S~\ref{sec5} we show the constraints on helium which allow to reproduce the hotter Grundahl jumps and in  \S~\ref{sec6} we discuss and summarize the results.

\begin{figure}
\label{F_directions}
\centering{
\vskip-35pt
\includegraphics[width=8.5cm]{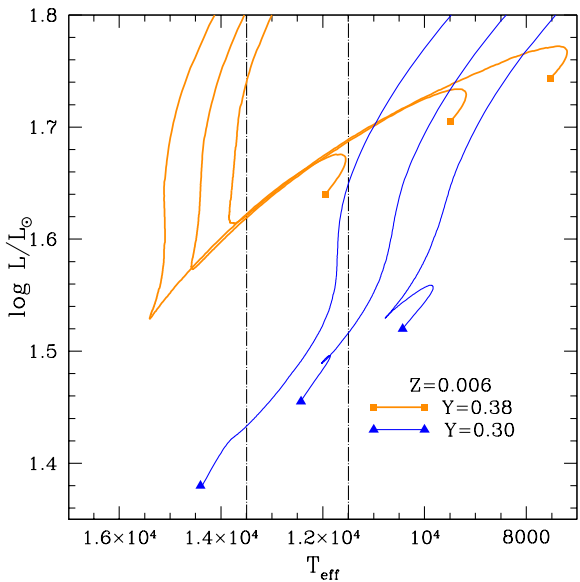}
\vskip -60pt
\caption{Comparison of the morphology of two different sets of stellar evolution tracks in the theoretical HR diagram. The ZAHB points are identified by squares for Y=0.38 and triangles for Y=0.30. The standard and hotter G--jumps are marked by dotted black lines. From hotter to cooler ZAHB location, the masses of the tracks are 0.53, 0.54 and 0.55\msun for both Y values. These tracks do not include helium settling, to display the prime effect of the initial helium content in the evolution. }    }
\end{figure}

\begin{figure}
 \label{figure1}
\centering{
\vskip-50pt
\includegraphics[width=9.7cm]{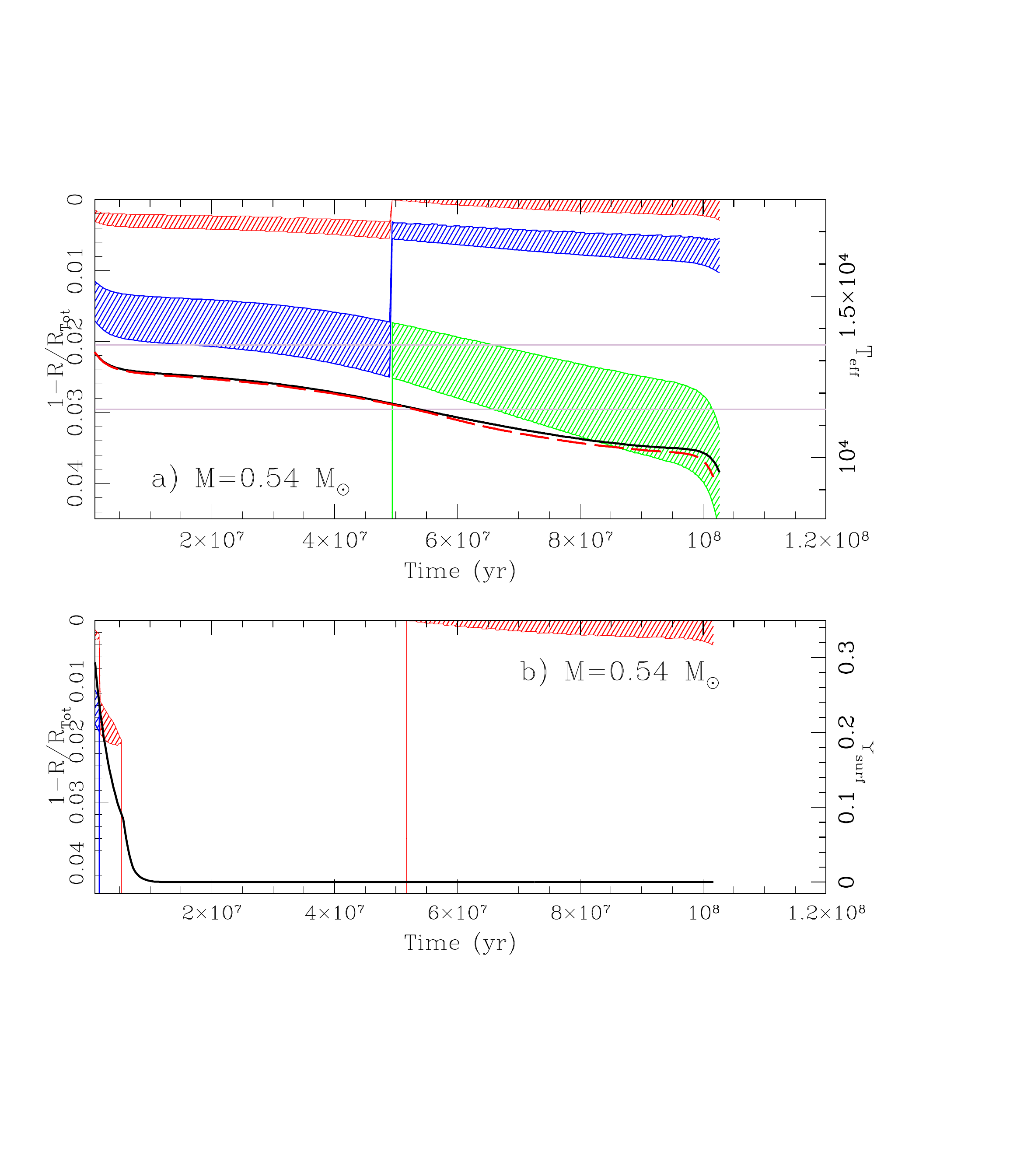}
\vskip -50pt
\caption{Time evolution of the HB convective regions for a track with Z=0.006, Y=0.28 and M=0.54\msun. Panel a (top) shows  the standard evolution, panel b (bottom) shows the evolution including He--diffusion. The dashed regions represent the location of convection zones, in terms of fractional radius (values going from 0 in the atmosphere to 1 in the center). The colors are red, blue and green in order, starting from the most external convective zone. The colors may change along the evolution if one or more specific convective regions appear or vanish when \teff\ evolves.
The \teff\ as a function of time (scale to the right of panel a) is shown for the standard track (black full line) and the track including diffusion (red dashed). Two horizontal lines signal the standard G--jump  (11,500\,K) and the anomalous G--jump \teff=13,500\,K. In panel b, the black line shows the evolution of \ysurf\ (scale on the right) in the track with diffusion.
The track evolves from high to low \teff, so the standard track has initially on top the He--I partial ionization region, and below the He--II region. At \teff$<$11,500\,K the partial H--ionization convection zone appears.
In panel b, we see that the two He-convection regions  disappear very soon, as helium settles in the layers where it would be partially  ionized. Surface turbulence reappears suddenly below the standard G--jump, where hydrogen is no longer fully ionized. The atmospheres along this whole \teff\ range have very small or negligible helium.}    }
\end{figure}

\section{The clusters NGC\,6388 and NGC\,6441: similarities and differences}
\label{sec2}
NGC\,6388 and NGC\,6441 host a very extended HB \citep{rich1997} in spite of their high metallicity, [Fe/H]$\sim$--0.50  \citep{carretta2007,origlia2008}; GC with this metallicity typically have only a red HB. The fraction of stars populating the red HB is the most prominent \citep[about 87\% in NGC\,6388, and about 91\% in NGC\,6441, ][]{bellini2013}. Another very important anomaly resides in the very long periods of their RR\,Lyr variables \citep[the average period of a,b type RR\,Lyr is 0.71\,d in NGC6388, and 0.76\,d in NGC\,6441][]{pritzl2000}, even longer than the typical periods of the Oosterhoff type II clusters, which are generally metal poor. The peculiar luminosity of the HB at the RR\,Lyr location is a clear indication of the presence of helium rich stars in the RR\,Lyr region, although this explanation was controversial until  \cite{caloidantona2007} modeled the whole HB of NGC\,6441 in the HST bands F439W and F555W observed by \cite{piotto2002}, showing that the helium enriched stars are not limited to the small percentage of the blue HB \citep{busso2007}, but are present also in the red HB, justifying its thickness in magnitude. In \cite{caloidantona2007} the percentage of stars having a standard--Y was determined as a mere 38\%  in NGC\,6441. A similar analysis was presented by \cite{dc2008} for the HB in NGC\,6388. In both clusters, the extension in color of the HB was attributed to the property of high--Y, high  metallicity HB models, which, for a  small range of mass, undergo long excursions from low to high \teff\ \citep{sweigartgross1976}.

Despite the fact that these clusters must host a significant fraction of very helium-rich stars, \cite{bellini2013} have shown that they must differ also in some other chemical aspects. In the color m$_{\rm F390W}$--m$_{\rm F814W}$\footnote{remember that the band F390W is influenced by CNO abundances.}, NGC\,6441 shows a split main sequence, populated by 65\% on its red side \citep{bellini2013}. The analysis by Bellini et al. shows that the helium difference between these two main sequences is $\Delta$Y$\sim 0.06$ and suggests that the helium distribution is bimodal, with about one third of NGC\,6441 stars clustering around very high helium (Y=0.35--0.36), and the remaining stars, including the first generation group, having on average Y$\sim$0.30. The distribution may be multimodal, similarly to what we have observed in NGC\,2808 \citep{dantona2005, piotto2007}, although either the presence of differential reddening, other observational peculiarities, or the presence of a strongly diluted second generation, in which the helium content merges with the initial value \citep[see, e.g.][for the case of NGC\,2808]{dantona2016}  do not allow to isolate independently the ``first generation" sequence. 

The visual color magnitude diagram of NGC\,6388 reveals a dim sub giant branch, that has been interpreted with a CNO enriched population including 22$\pm$2\% of the total number of cluster stars  \citep{piotto2012}. In other clusters with split sub giant branch, the CNO variation has been confirmed from direct spectroscopic measurements \citep{yong2009, yong2015, marino2012m22, lim2015calcium}. In contrast, there is no evidence of a CNO enriched population in NGC\,6441 \citep{bellini2013}. On the other hand, the main sequence color spread in  m$_{\rm F606W}$--m$_{\rm F814W}$ of the two clusters is similar, implying similar helium total variation, as also suggested by the HB color and magnitude distribution. 

The situation is very complex, and a more general discussion on the HB properties of the Treasury Program clusters will be presented in a separate paper. We devote this first study to the presentation of the models and to constrain the value of the helium content of the stars crossing the CM diagram from the red to the blue side, with detailed comparison with the location of the G--jump.

\section{HB models including sedimentation}
\label{sec3}
The HB models have been  computed with the ATON code, whose main inputs concerning opacities, convection model ---including overshooting --- and equation of state are reported in  \cite{ventura2008code}. The most recent updates in the models follow those presented in \cite{tailo2015,tailo2016}, apart from a refined treatment of diffusion which we discuss in this section.
Complementary isochrones for the evolution from the main sequence up to the tip of the red giant branch, with the same compositions of the HB models, are computed.
The HB evolution starts from the zero age horizontal branch (ZAHB), where the helium-core mass is fixed by previous evolution up to the helium flash. We follow the evolution of each model up to the exhaustion of helium in the core.

In this work, we adopt a metallicity in mass fraction Z=0.006, with [$\alpha$/Fe]=0.40 based on the abundance measurements by \cite{carretta2007} for  NGC 6388  and by \cite{origlia2008} for NGC 6441. 
The helium mass fraction is set to Y=0.25, 0.28, 0.30, 0.32, 0.35 and 0.38. We compute both standard models, and models including helium diffusion, using the formulation given by \cite{thoul1994}. The routine self consistently  computes the effects of gravitational and thermal settling of helium and of the `average' metal with respect to hydrogen, in all the mesh-points of the structure palced within an outer and inner boundary. The inner boundary is placed where hydrogen is present. The outer boundary is chosen as  described in the following paragraphs. The code computes the coefficients described in equation 41 of \cite{thoul1994}  to obtain the diffusion velocities for each of the components considered. Then, the equation regulating the rate of change of the elements mass fraction \citep[equation 40 in][]{thoul1994} is solved following the scheme adopted in \cite{ibenmacd1985}. 

The treatment of diffusion --even the simple diffusion of helium-- is a challenging problem. Note that  the diffusion velocities rapidly decrease as density and temperature increase. Surface settling of helium is very fast in the optical atmosphere, unless some source of turbulence is present, so that diffusion acts only below the turbulent region, Thus, the global effect of diffusion is a  by-product of the  residual turbulence in the outer envelope \citep{michaud2007, michaud2008}.
A common way to deal with this problem is to assume as outer boundary of the diffusion computation  a fixed mass fraction \Mturb, above which it is assumed that  turbulence prevents diffusion. 
Values of \Mturb$\sim $10$^{-3}-10^{-8}$ of the total mass are quoted in relevant literature, for main sequence low mass stars with deep convective envelopes. If the inner boundary of the convective region is inner than \Mturb, diffusion operates from the inner boundary of convection.

\begin{figure}
\label{F1}
\centering{
\vskip-50pt
\includegraphics[width=9.7cm]{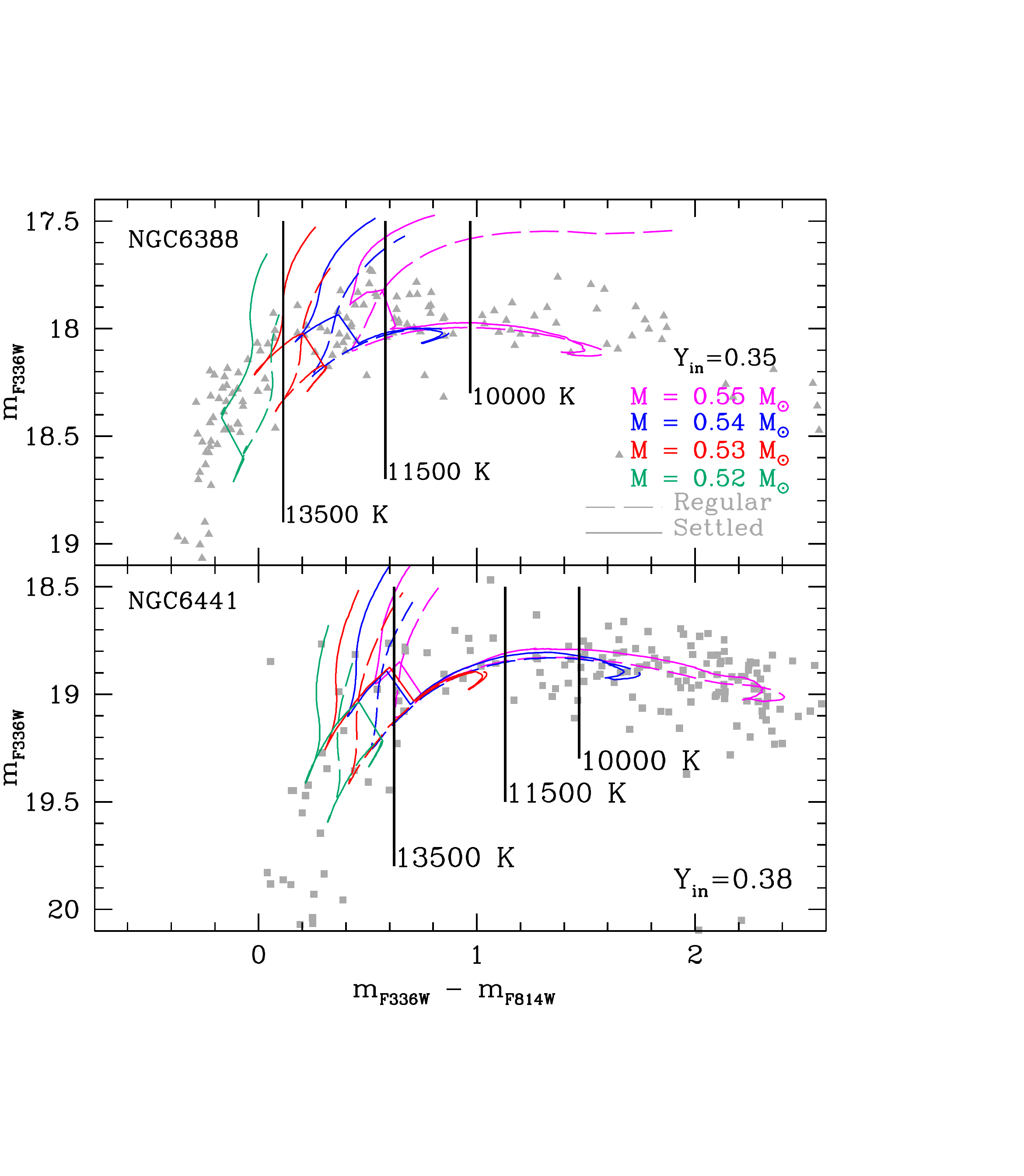}
\vskip -50pt
\caption{The data for the blue side of the HB of NGC\,6441 (bottom panel, squares) and NGC 6388 (top panel, triangles) are shown in the m$_{\rm F336W}$\ versus \col\ plane. Magnitudes and colors are not corrected for absorption and reddening, but the vertical lines indicating \teff=10,000\,K and the two G--jumps at 11,500 and 13,500\,K serve as a reference for the \teff. We superimpose the tracks of four masses (0.52, 0.53, 0.54 and 0.55\msun, from left to right).  The standard tracks without diffusion (dashed) and those with He--diffusion (full lines) are shown. The discontinuity in the tracks with diffusion are due to the switch from the correlations magnitude--luminosity of the model metallicity to those with [Fe/H]=0, to empirically account for the metal levitation in the stable atmosphere, when helium is fully depleted . }    }
\end{figure}

\begin{figure*}
\label{interiors}
\centering{
\vskip-50pt
\includegraphics[width=20cm]{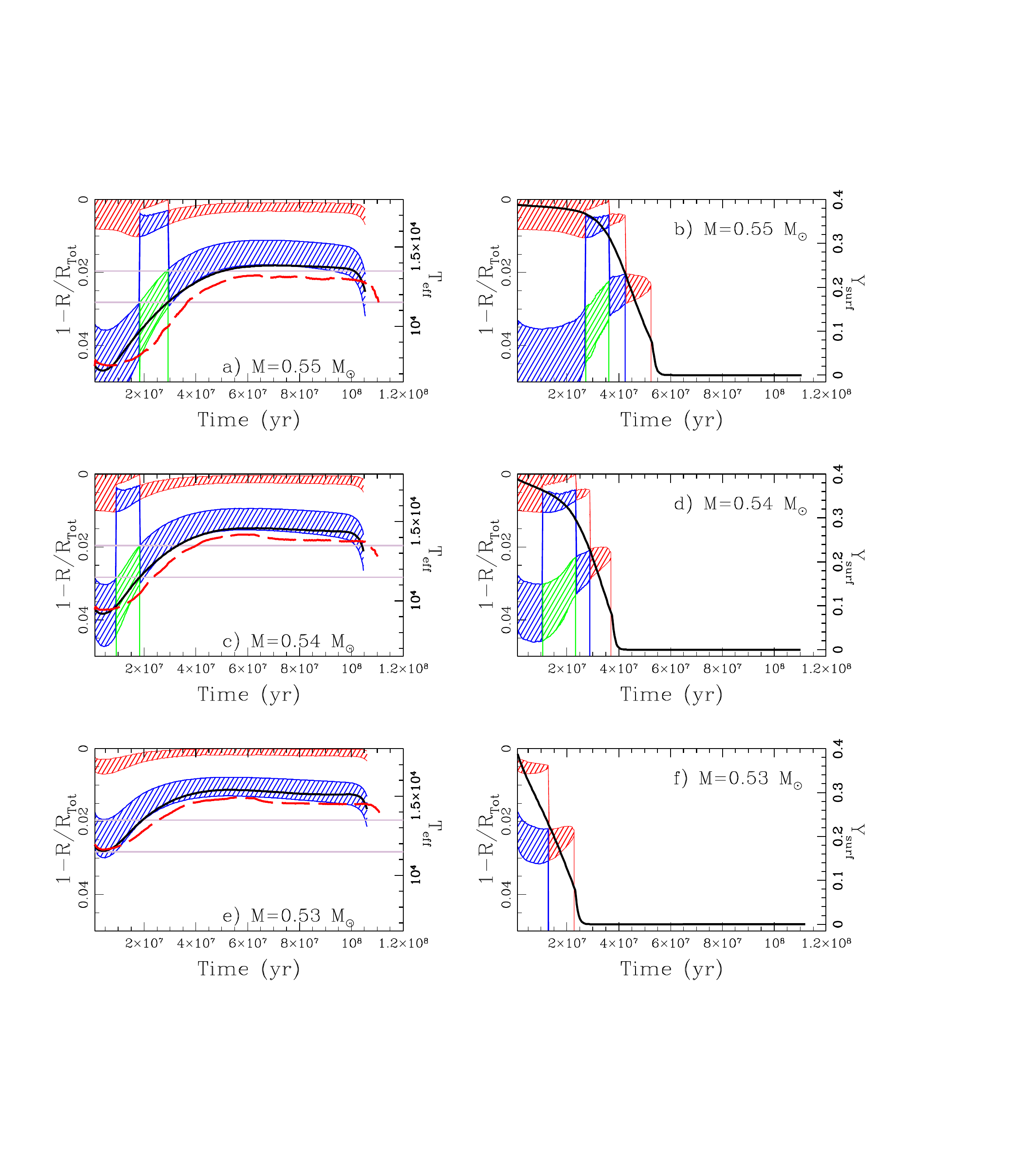}
\vskip -100pt
\caption{The display setup is similar to Fig. 2. Evolution of the convective regions for models of Y=0.38 and M=0.55\msun\ (panels a and b), 0.54\msun\ (c, d) and 0.53\msun\ (e, f). In the left panels are the standard tracks, where the He--I convective region is always present, although it does not reach the surface.
The tracks on the right include He--diffusion, which modifies the convective He regions. All three stars cross the \teff\ region 
in between the G--jump preserving at least the He--II convective region, and become fully radiative at \teff$\sim$13,500\,K. 
  }    }
\end{figure*}

\begin{figure*}
\label{convY38}
\centering{
\vskip-50pt
\includegraphics[width=20cm]{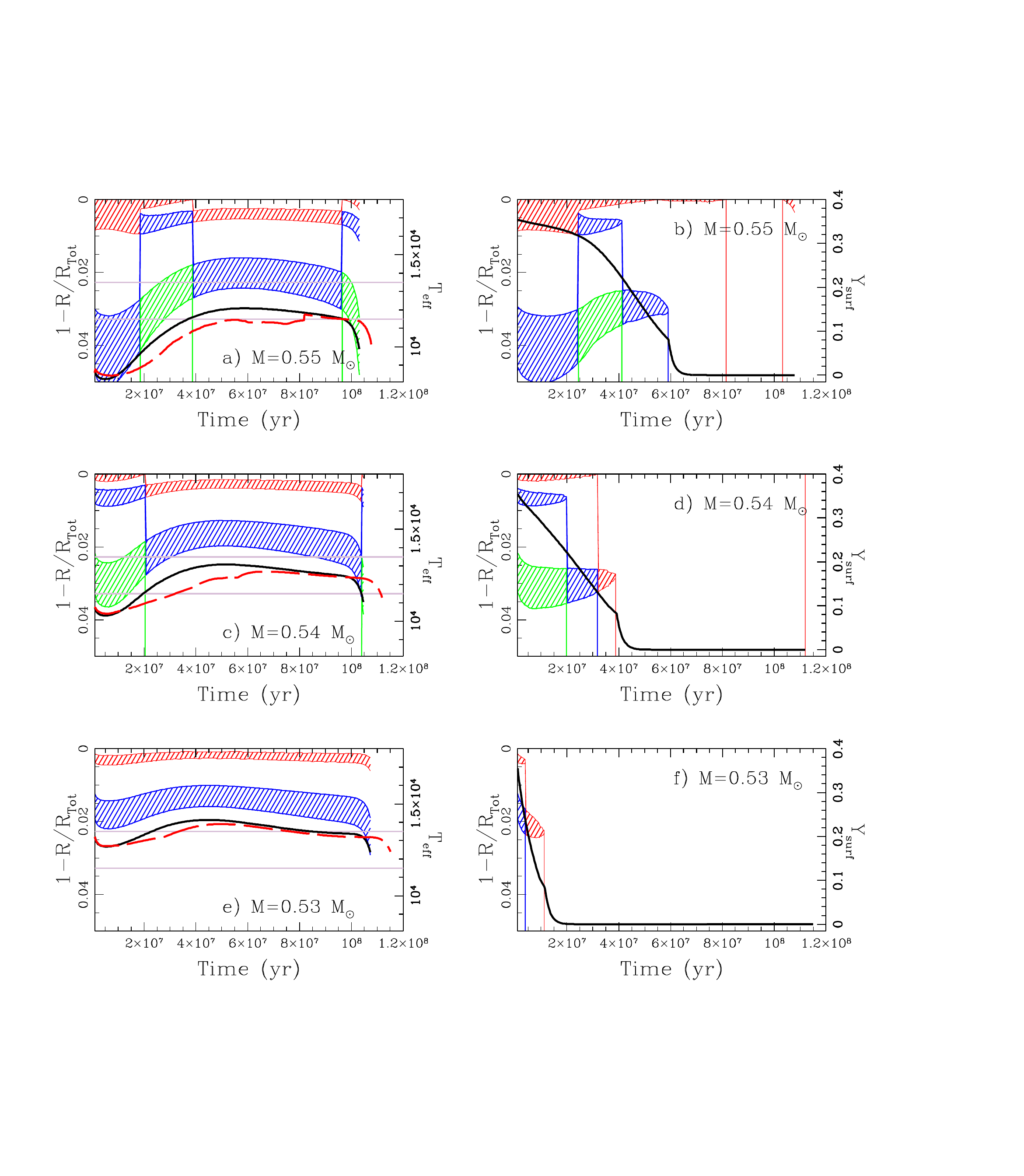}
\vskip -100pt
\caption{ Similar to Fig. 4, but  for models with Y=0.35. The \teff\ excursions are shorter, and intermediate cases are present in the evolution from convective to radiative structures.
The 0.55\msun\ evolves from smaller \teff until the border of the standard G--jump, and \ysurf\  slowly decreases along the evolution, until its depletion suddenly accelerates when the He--II convection disappears. The 0.54\msun\ evolution shows the case in which the survival of the He--II convection zone  postpones the G--jump to \teff $\simgt$11,500\,K. The 0.53\msun\ evolution starts at \teff$>$10,000\,K, and achieves full depletion of helium at \teff$\sim$13,000\,K. }    }
\end{figure*}

For HB models, there are several important studies in the recent literature. \cite{michaud2011} include self consistently both atomic diffusion and radiative acceleration. They find that the abundance anomalies observed in HB stars atmospheres at \teff\ above the G--jump are compatible with the model results, when they assume an envelope mixed mass of $\sim 10^{-7}$\msun.  No hypothesis is made about the physical reasons for the mixing mechanism. The same models predict {\it too large} helium abundances with respect to the observations in the range 12,000$\simlt$\teff$\simlt$17,000\,K \citep[see also][]{moehler2014}. 
 \cite{leblanc2009, leblanc2010} have shown that stratified model atmospheres, in which atomic diffusion and levitation are both included solve the long-standing problem of computing appropriate gravities for the hot HB stars, and this seems in contrast with a fully mixed region \citep{moehler2014}. 
Nevertheless, these same stratified models predict higher than observed abundances for the elements affected by radiative levitation. Thus these results are in part contradictory and much work is needed to arrive to a self-consistent solution. 

Concerning our models, we concentrate on studying the interaction between the timescale of \teff\ change with time along the evolution and the timescale of helium diffusion. We wish to explore what happens when these two timescales are comparable. 
As observations and models have already shown that the G--jump transition is concomitant to the appearance --or disappearence-- of the H--convection region, we have to follow as completely as possible the evolution of the convective regions and the transition from convective to radiative envelopes. 
The computations will determine when and at which \teff\ the atmosphere becomes free of any kind of turbulence, and radiation pressure on the lines allows metals to levitate, producing the G--jump.  
The variation with \teff\ of the convective regions can not be described by choosing a boundary for the turbulence too much inside the star. The external layers may be convective due to the partial hydrogen ionization region for \teff$\leq$11,500\,K, but the He\,I and He\,II partial ionization regions may be present in deeper layers (see Fig.\,2, top panel), and they emerge into the most external envelope respectively at \teff$\sim$15,000\,K and $\sim$30,000\,K (see, e.g., figure 9 in paper VII and references therein). These convective regions should play a role in the diffusion modeling.

We first examined models in which  diffusion is efficient {\it below the inner boundary of the most external  convective region (either He\,I or H), or from below a very small mass fraction (\Mturb=10$^{-10}$ of the total mass) if there is no outer convection.}
Sedimentation of helium in the atmosphere results too fast, and is not consistent with the measured (small) abundances of \ysurf\ in the HB stars in NGC~6752 at  \teff$<$12,000\,K  \citep{monibidin2007}. In order to fit the observed values,  we may counteract the fast helium settling with a parametric mass loss along the evolution. This approach is successful, but it is not satisfactory, because the mass loss rate necessary to fit the data vary with \teff. Those required for models at \teff$<$11,500\,K are $\sim$10 times higher than  at \teff$>$11,500\,K.  The reason is straightforward. In models cooler than the G--jump, our assumption implies that diffusion operates below the (small) surface convective region due to partial H--ionization. On the other hand, for models at \teff\ higher than the G--jump, it operates from below the He\,I--convection region, where the diffusion velocities are low. 

We moved to a different approximation. We still fix the external boundary for the diffusion computation at \Mturb=10$^{-10}$ of the total mass, if both H-- and He\,I--convection layers are not present or do not extend beyond \Mturb. This choice helps to make stable the solution of the diffusion equations. In all other cases, 
{\it we assume that the whole external envelope is turbulent down to the boundary below the He\,II convective region, independently from the presence or not of the H--convection layer}. The diffusion velocities below this boundary are smaller than below the plain H--convection layer, so diffusion less effective also at \teff\ below the G--jump. This choice of the external diffusion boundary proved to be more satisfactory, as it requires only a very small mass loss rate (of the order of the solar mass loss rate) to allow the survival of the small but measurable helium content in the stars hotter than the G--jump, in agreement with the data of NGC~6752 (Tailo et al. 2017a, in preparation). For these reasons we adopt this approach in the present work.

We use bolometric corrections and color transformations available from \cite{castellikurucz2004}, for the metallicity Z=0.006 and for standard helium content. The models do not include atmospheric radiative levitation so they are not fully self--consistent. We simulate its effect by shifting the specific magnitude computation to the tables of [Fe/H]=0.3 transformations {\it as soon as the helium surface content of the model drops below \ysurf=10$^{-3}$}. In other words, although we use a rough approximation\footnote{ \cite{leblanc2010} show that  model atmospheres in which the metals are homogeneously increased account only in part for the offset  of the G--jump in the \cite{grundahl1999} Str\"omgren color magnitude diagram. Computation of stratified model atmospheres is necessary to account for the whole G--jump. }, we link the occurrence of metal levitation to the evolution of the surface helium abundance experienced by our models, which is a good proxy of the surface stellar turbulence in the \teff\ range of interest. 

\subsection{The necessity of full evolutionary computation to model the role of helium in the models}

One immediate result of our test computation of the HB evolution models including diffusion is the following. 
Convection due to the high opacity in the hydrogen partial ionization zone disappears at the onset of full ionization, at \teff=11,500\,K, which establishes a very well defined  convection boundary (and consequently is a strong observational marker).
The behavior of the He\,I and He\,II partial ionization regions is much more awkward: their survival also depends on the {\it presence} of helium in the layers with the appropriate temperatures. {\it If helium is fully settled in these layers, one or both of these convective regions may disappear in the course of evolution}. 

During their HB lifetime, stars do not remain at a fixed \teff, so the location and extension of their convection zones will change because the \teff\ changes. The variation of the convective layers depends also on the timescale on which this \teff\ evolution occurs compared to the diffusion timescale.
This effect is schematically described in Fig.\,1, where we compare two different sets of evolutionary tracks whose ZAHB locations (solid squares and triangles) are in proximity of the G--jump region. The two sets refer to models with initial helium (\yin) equal to 0.30 and 0.38, for masses 0.53, 0.54 and 0.55\Msun. The paths of the plotted tracks are different. Tracks with lower \yin\ evolve from high to low \teff, whereas those with higher \yin\ cross the G--jump boundary (11,500\,K) from low to high \teff.

\subsubsection{Standard evolution: reaching the G--jump from a larger \teff}
{\it If helium diffusion is accounted for and the evolution proceeds from an initial \teff$>$11,500\,K, helium settles  to layers where it is fully ionized, and the convection regions disappear for any values of  \yin\ (standard or higher than standard)}. In the stable atmosphere metals levitate, until the evolving star crosses the boundary of 11,500\,K and the onset of H--convection destroys the atmospheric radiative stability. 
In general, the tracks in the G--jump region evolve from hotter to cooler \teff, for both standard and enhanced Y, at the low metallicity typical of most GCs. 

This is the `standard' case, and we depict it by looking at the evolution of the convective regions  (Fig.\,2) for the 0.54\msun\ with Y=0.28 and Z=0.006. This track crosses the \teff\ range from $\sim$13,000\,K  to $<$10,000\,K. During the whole standard (no diffusion) evolution the He--partial ionization regions in the inner layers are preserved (top panel), but these regions disappear very soon including the effect of diffusion (bottom panel). 
In models with important mass loss rates (${\rm dM/dt}\sim 10^{-8}$M$_\odot$/yr), or having a bigger \Mturb, helium settling will be slower than shown in Fig.\,2, but the helium convective region will be quenched before the star evolves through the standard G--jump, as the \teff\ decrease with time takes place over several tens of million years, and models must be consistent with the small \ysurf\ at \teff$>$11,500\,K in the observations\citep{monibidin2007}.

The stable atmosphere allows metals levitation, so that the G--jump appears strictly below \teff=11,500\,K. This kind of evolution is standard for all tracks with Y$\simlt$0.34 at low metallicity. This justifies the unique location of the G--jump at the same \teff\ in all GCs \citep{brown2016}, in spite of possible differences in the specific \yin\ of stars populating the \teff's close to the G--jump in clusters having different histories of helium enrichment in second generation stars. 

\subsubsection{Crossing the G--jump evolving from a smaller \teff}
The peculiar case of the clusters under examination is illustrated in Fig.\,3, where  we show the Treasury program data for NGC~6388 and NGC~6441.  The two observational G--jumps  are indicated by vertical lines at \teff=11,500\,K and 13,500\,K \citep[see the discussion in][]{brown2016}. Superimposed are tracks for Z=0.006 and Y=0.35 (top) or Y=0.38 (bottom), both standard (dashed) and including diffusion (full lines).  Diffusion modifies the \teff\ excursions of the tracks, which become shorter. The onset of the effect of radiative metal levitation (the change of bolometric corrections, leading to the sudden jumps in color and magnitude) is included when the surface Y becomes $<$0.001.

\begin{figure}
 \label{figgeneral}
\centering{
\vskip-50pt
\includegraphics[width=9.7cm]{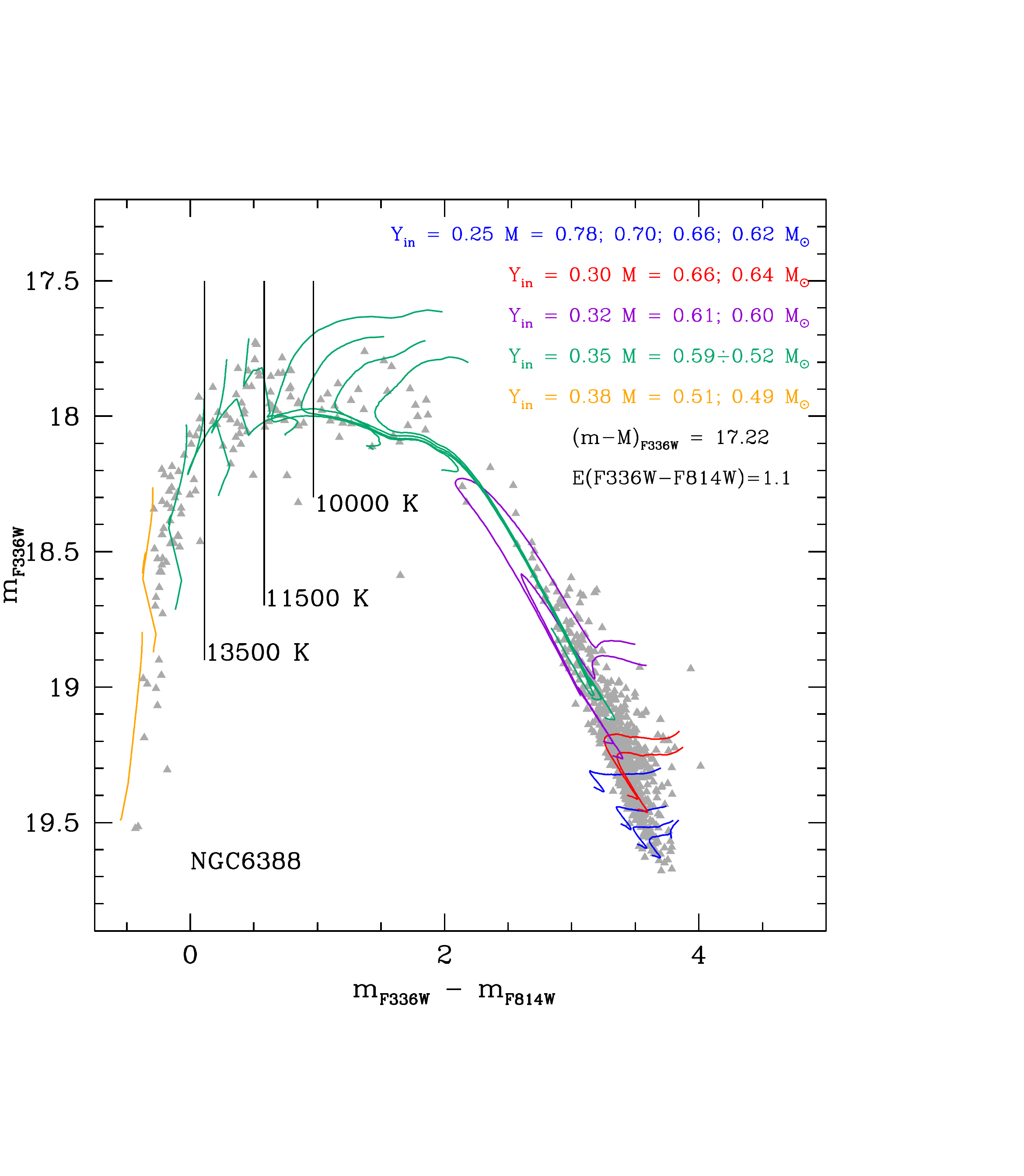}
\vskip -50pt
\caption{The most relevant evolutionary tracks in the simulation of the HB of NGC\,6388, superimposed to the data. Color of the tracks corresponds to the initial Y, according to the figure label. The distance modulus and color reddening attributed to the tracks are labelled. }    }
\end{figure}

\begin{figure*}
\label{global}
\centering{
\vskip-100pt
\includegraphics[width=20.5cm]{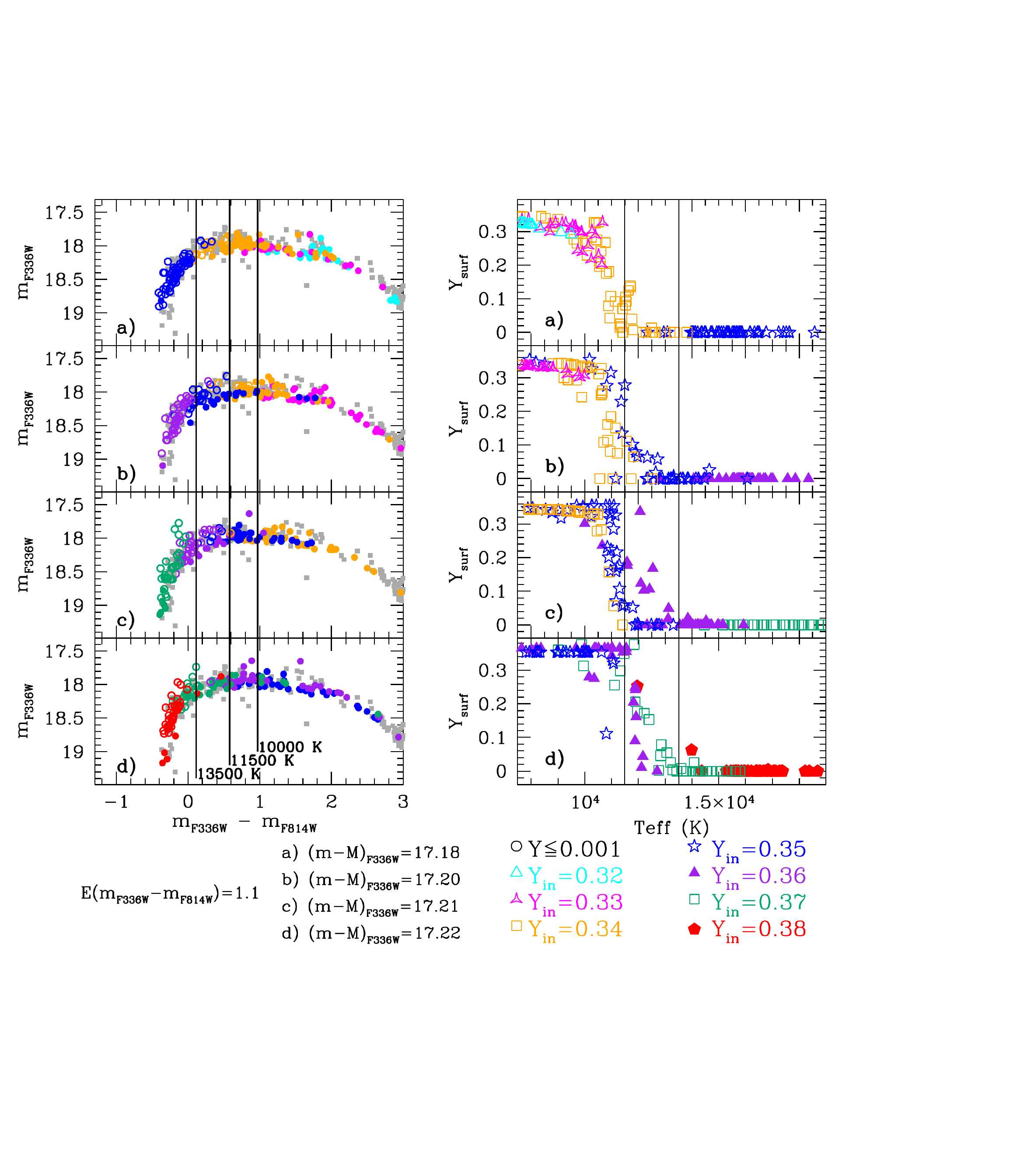}
\vskip -120pt
\caption{Left panels: simulations (dots) are superimposed to the blue HB data (grey squares) of the cluster NGC\,6388. Different input helium distributions, increasing from panel a to d, are assumed along the HB. The different initial Y of the dots are highlighted with the colors labelled at the bottom of the figure. The distance modulus is labelled too. \teff=10,000, 11,500 and 13,500K  are marked for reference. The open circles in the left panes are the points where diffusion has reduced \ysurf\ below 10$^{-3}$. Note that  the number of stars in which helium not full settled at 13,500$<$\teff$<$11,500\,K increases when Y increases, thanks to the longer loops of the tracks with higher Y. This is shown in details in 
the right panels, where we show the \ysurf\ versus \teff\ for the four simulations. Here the different helium contents are marked with different symbols, labelled at the bottom.  The 11,500 and 13,500 G--jump boundaries are marked. }    
}
\end{figure*}

Apart from the smallest mass (0.52\msun), {\it the tracks evolve from low to high \teff.} 
If the evolution starts from \teff$<$11,500\,K, the He--abundance in the envelope may remain large enough that the He convection zones may survive, possibly until this first He ionization is complete, formally at a \teff\ as large as $\simeq$15,000\,K.  The stability of the atmosphere and the occurrence of radiative levitation will then occur only at a  \teff\ intermediate between the standard G--jump (11,500\,K) and the \teff\ of full He--ionization. 

We see that the hotter track (0.52\Msun), evolving from high to lower \teff, shows a prompt effect of diffusion. The other tracks show loops in the HR diagram. The long loops are due to the very high Y and the relatively high metallicity of the models, as shown in the seminal paper by \cite{sweigartgross1976}. The most interesting cases are the 0.54 and 0.55\msun\ evolutions, which proceed along long loops from \teff\ below the G--jump.  Indeed the 0.54\msun\ track shows that the G--jump occurs at larger \teff\ than the standard models. 
The precise \teff\ at which the stellar envelope attains radiative equilibrium depends on the competition between the He--diffusion timescale of the models and the timescale of \teff\ evolution. The larger is Y, the faster is the evolution through the G--jump region. \cite{brown2016} analysis shows that NGC\,6441 has only the hotter G--jump, while NGC\,6388 may have also a (smaller) G--jump at the standard \teff. This may hint that some of the stars in NGC\,6388 are evolving along tracks which become fully radiative at the standard \teff. We show that this may depend on the precise Y of the stars evolving in this \teff\ range.

In the following figures, we show the difference between the convection region in models with and without gravitational settling for Y=0.38 (Fig.\,4) and Y=0.35 (Fig.\,5). 
We depict the evolution of the outer convection zones for three tracks in the standard (left panels) and diffusion case (right panels). The full black line and red dashed line in the left figures represent the \teff\ evolution with time (right side scale). On the other hand, the black lines of the right panels describe the \ysurf\ evolution with time of the diffusive models (scale on the right).
We see that diffusion drastically modifies the He--convective region. 
In Fig. 4 (case Y=0.38) the evolutionary loops are very extended, and the fully radiative structure is achieved at \teff$\sim$13,500\,K in all the three tracks. These models will produce a sharp G--jump at the required \teff.
We allow diffusion below the He-I partial ionization layer, but notice that the He-II driven convective region, which is interior to the 2\% of the stellar radius, also plays a role and acts to slow down He--diffusion efficiently (see later the discussion on the observables).
This shows that the diffusion problem is far to be fully explored, and these models are only first steps towards a full understanding of the different physical inputs.

In Fig. 5 (case Y=0.35), the top panels (a and b) represent the 0.55\msun\ evolution, which starts at \teff$<$10,000\,K and evolves rapidly towards larger \teff. The track excursion ends more or less at the standard G--jump. The 0.54\msun\ track (panels c and d) evolves through the standard G--jump. It preserves both the He--I convection region and a finite \ysurf\ until \teff$\sim$12,000\,K. The 0.53\Msun\ track (panels e and f) starts in between the two G--jumps and loses the convective regions at \teff$\sim$13,000K. 

\begin{figure}
\label{6388}
\centering{
\vskip-30pt
\includegraphics[width=9.7cm]{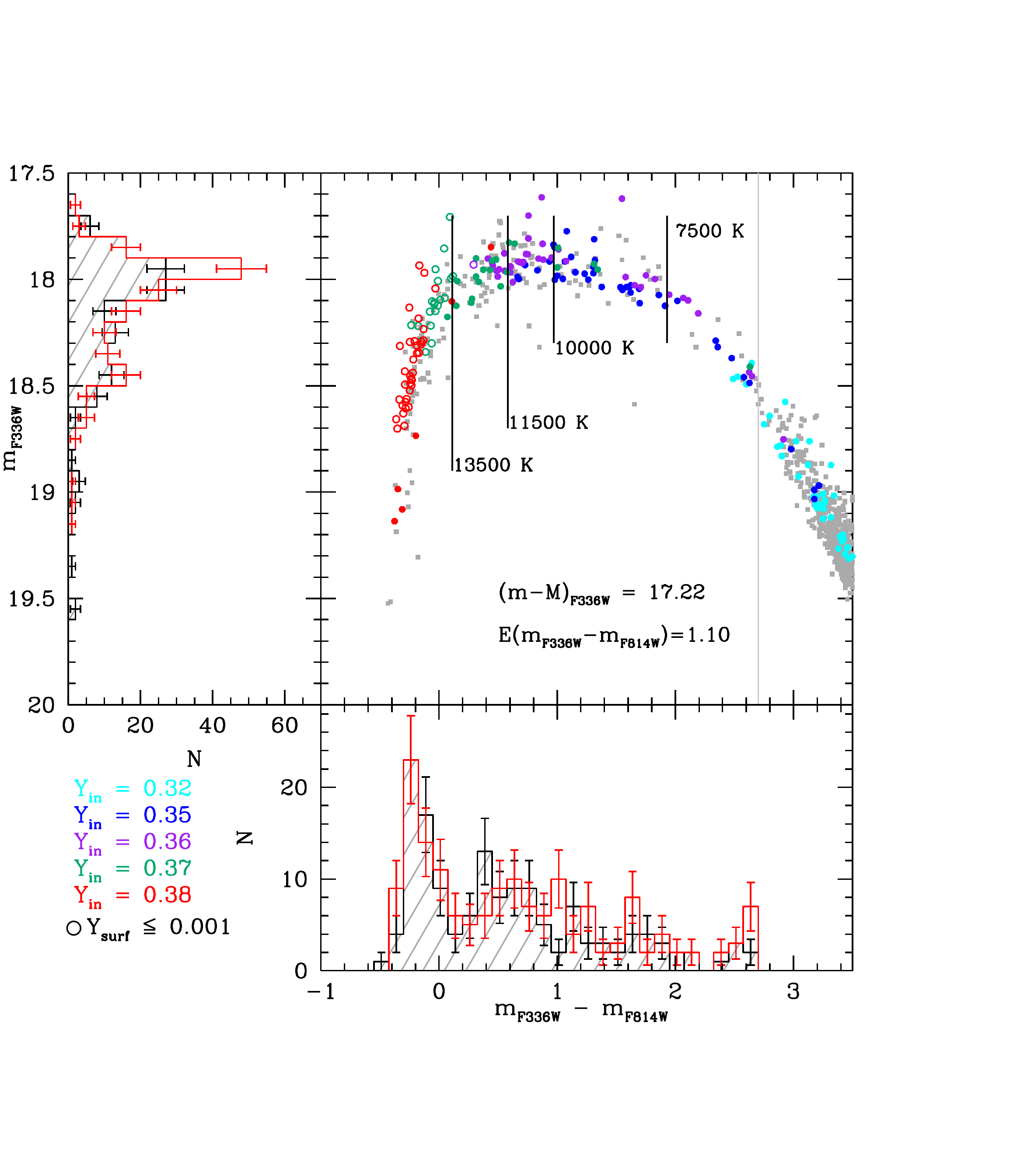}
\vskip -40pt
\caption{Full simulation for the HB data of the cluster NGC\,6388. The distance modulus and reddening used are labelled in the figure. The histograms in the bottom panel show the color distribution of stars with \col<2.7 whereas the ones in the side panel represent the number of stars versus m$_{\rm F336W}$. (black with grey shading: stellar counts; red:  simulation). The observations are shown as grey squares, and the simulated points are color-coded according to labels in the figure. Open dots represents the simulated stars where the helium is fully settled The line at \col=2.7 shows where the histogram stops, but the simulation extends to slightly redder colors. }    }
\end{figure}

\begin{figure}
\label{6441}
\centering{
\vskip-30pt
\includegraphics[width=9.7cm]{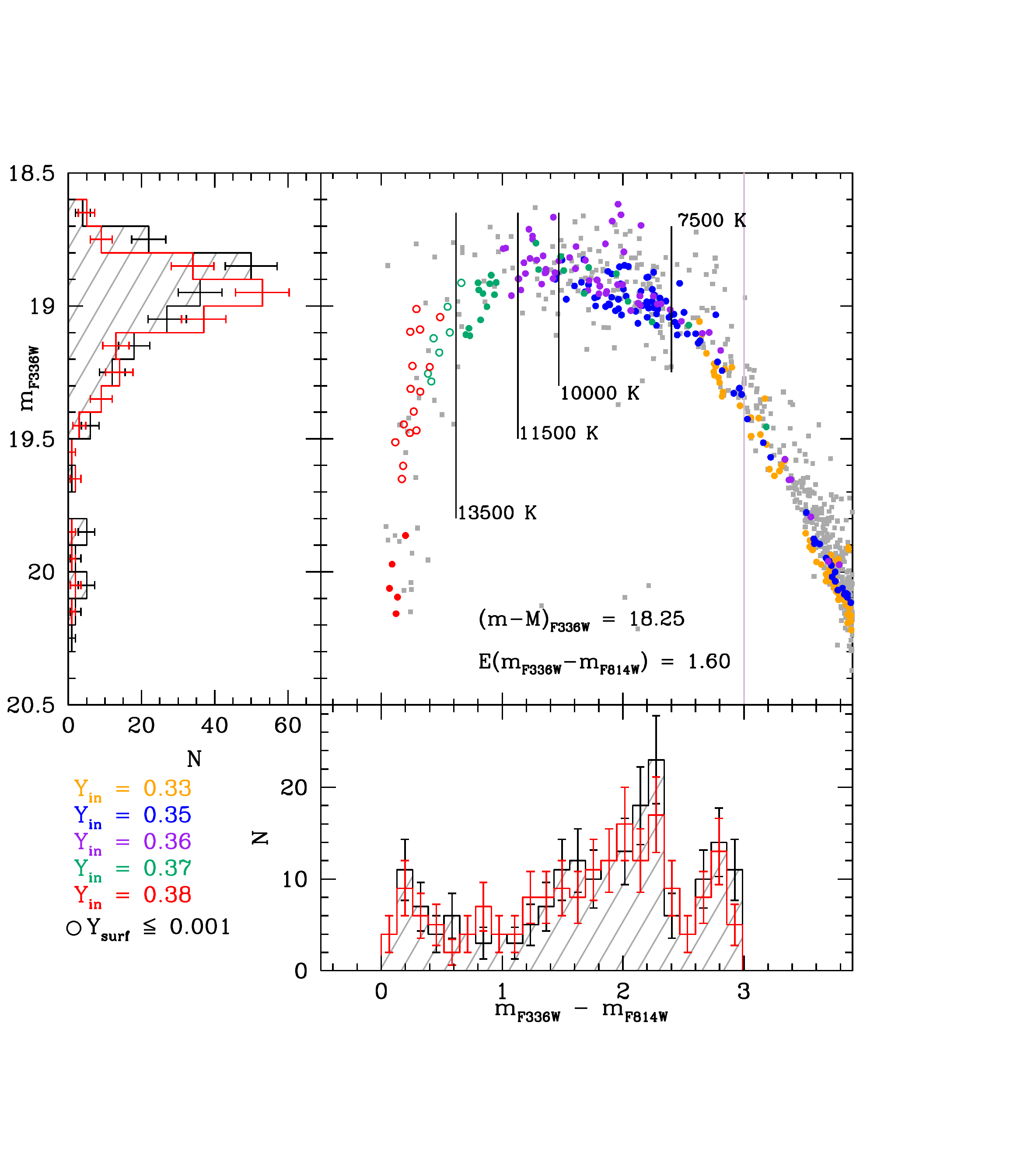}
\vskip -40pt
\caption{The same as in Fig. 8 but for NGC\,6441. Here in this figure the histogram stops at \col=3.0   }    }
\end{figure}

\section{Simulations for the blue HB side}
\label{sec4}

We compute simulations of the HB data from the Treasury program using the new diffusion tracks database. This allows us to check the working hypothesis that the interplay between the timescale of diffusion and the timescale of the evolution beyond the standard \teff\ of the G--jump is the dominant physical reason which shifts the G--jump location  to \teff$\sim$13,500\,K in these peculiar clusters.  

We show in Fig.\.6 the whole HB of NGC\,6388 in the plane m$_{\rm F336W}$ versus \col. The UV band data come from the Treasury HST Survey \citep{piotto2015} while the m$_{F814W}$\ data come from the ACS Survey of GCs \citep{sarajedini2007}.
Appropriate tracks from the database ---shifted to apparent magnitudes and colors--- are superimposed on the data. As already well known from previous work, different helium contents are needed to reproduce the color--magnitude diagram. The clump must include stars with Y as large as Y=0.32. In addition, values of Y$>$0.34 are needed to `cross' the diagram from low to large \teff, at the appropriate magnitude \citep[as shown by][ this is necessary also to reproduce the RR\,Lyr anomalous long periods and high luminosities]{caloidantona2007}.
The latter authors presented example simulations of the optical HB data of these clusters, while we now concentrate on the blue side only. As the helium content of the stars which evolve from the red to the blue side is not known from first principle, we use different simulations to understand which are the plausible choices.

The synthetic HBs are built by  basically following the prescriptions described by \cite{caloidantona2007} \citep[see also ][]{tailo2016}. We fix the age of the cluster (typically at 12\,Gyr) and subdivide the stars into groups having different helium content. We derive the mass \Mrg\ evolving in the red giant phase for the set (age, Z, Y) of each group, and assume an average mass loss value $\delta$M, which is randomly varied assuming a gaussian deviation $\sigma(\delta$M). We thus have an HB mass  M$_{\rm HB}$=\Mrg--$\delta$M for each group of Y values. The average mass loss is adjusted to achieve the M$_{\rm HB}$\ masses which describe the observational values. In principle, the variation of M$_{\rm HB}$\  may be due solely to the decrease of \Mrg\  with increasing Y, but previous studies have shown that some additional mass loss may be needed \citep[see, e.g., the discussion in][]{dantona2013}. We take care to avoid discontinuities in the number counts and/or in the Y values. If the samples are similar in numbers, or monotonically varied with Y, the discontinuities obtained in the color distribution are a result of the temporal evolution of the tracks. Example cases of the chosen values are listed in Table\,1.

\subsection{The blue HB description in terms of models including diffusion}
\label{sec5}

In Fig.\,7 we present a simple test of our hypothesis concerning the role of helium in the shift of the G--jump. 
Considering only the HB part extending from the RR Lyr region to the Momany gap at \teff$\simeq$20000\,K, we show the influence of a different choice of the helium distribution for stars crossing the \teff\  range between the standard G--jump (11,500\,K) and the hotter G--jump location (13,500\,K). 
In the left panels of Fig.\,7 we show four simulations for the HB of NGC\,6388 (coloured dots), superimposed to the data (grey squares).  The helium mass fraction increases from the panel a (Y=0.34 in between the G--jumps) to panel d (Y=0.37). We slightly adjusted the value of mass loss to compensate for the variation of \Mrg\ induced by the increase in Y. This is done to obtain stars of grossly the same \teff's in all four simulations.
 The luminosity of the tracks increases with Y, so we adjust the distance modulus to fit the different simulations to the data. 
 The HB luminosity fit requires an increase by 0.04\,mag, and the average period of the RR~Lyr increases from 0.644\,d to 0.742\,d,  passing from the simulation of panel a to d.  
Considering that the mean period of a-b type RR\,Lyr is 0.71\,d in this cluster \citep{pritzl2001}, and the uncertainty in the metallicity, the agreement can be considered good for all the simulations. The determination of the helium content from the RR\,Lyr confirms that it must be large, in the range Y=0.32--0.36.

The right panels in Fig.\,7 show \ysurf\ in the different simulations. We see that in the panel a all the stars deplete the surface helium when they cross the standard G--jump: this indicates that Y=0.34 is not sufficient to provoke the shift of the main G--jump. 
In panels b and c, a fraction of the stars, with Y=0.35, shifts full He--settling to $\sim$12500\,K, and panel d shows that most stars preserve a non negligible helium content up to \teff$\simeq$13,000 - 13,500\,K if Y=0.37 in between the two G--jumps. 
These simulations show that  a stable atmosphere and radiative levitation of metals may occur at higher \teff\ than in standard clusters, as shown in \cite{brown2016}. Accordingly,  the effect of survival of the convection region described in \S\,\ref{sec3} for the tracks evolving from low to high \teff\ becomes efficient enough only when Y=0.35-0.37 in the G--jump \teff\ range.

The standard G--jump at \teff=11,500K is not evident in the data of NGC\,6441 (a small stellar sample is available), while it might be present in NGC\,6388 \citep{brown2016}. Anyway, panel c in Figure 7 suggests that a standard G--jump could also be present and coexist with the hotter G--jump in these clusters.

If the helium abundance in the \teff\ range 11,500--13,500\,K is not negligible, it could be revealed by appropriate observations. In spite of the very high {\it initial } Y, a fraction of stars should show  0$<$\ysurf$\leq$0.2, and should not display peculiarly large metal abundances.

\subsection{Two simulations for the blue HB}

The simulations exemplified in Figure 8 and 9 have helium distributions chosen according to panel d of Fig.\,7 for both NGC\,6388  NGC\,6441.  The data are shifted in distance modulus and dereddened according to the labels in the figures, to be superimposed to the simulation points. The comparison is made through the number versus magnitude m$_{F336W}$\ histograms, where the data are shown as dashed grey histograms in the left panel. 
The histograms represented in the bottom panel of both Figs.\,8 and 9 show the stellar counts versus color, limited to \col=2.7 and to \col=3.0 respectively. The simulation correctly describes the rapid excursion from the red to the blue side in both clusters, evident in the very low stellar counts at  \teff$\sim$7,500\,K. 
Different colors identify stars extracted in the different groups of helium, as labelled. 
The main inputs of the simulations are in Table\,1. We use four groups of Y uniformly spaced for the bluest part of the branch. The mass loss $\delta$M during the red giant evolution is assumed to be slightly higher for the Y=0.37 and 0.38 groups, to allow a better fit of the morphology in the chosen schematization. 
The number versus Y distribution is flat in the case of NGC\,6388, and it is monotonically decreasing for NGC\,6441, but no variations in the numbers are artificially introduced to reproduce the number {\it vs.}color  discontinuities. In particular, we see that the dip in the counts for NGC\,6388 at the right of the \teff=13,500\,K boundary is all contained within the same helium (Y=0.37) and mass loss group, so it reflects mainly the \teff\ versus time evolution of the tracks, the effect of helium settling and the onset of the G--jump. In the case of NGC\,6441 the situation is similar, but the scarcity of data does not allow a more stringent comparison. 

\begin{table}
	\centering
	\caption{ Number of simulated stars stars for different values of helium and mass loss.}
	\label{tab:example_table}
	\begin{tabular}{lrcclrc}
		\hline
\multicolumn{3}{c}{NGC 6441} && \multicolumn{3}{c}{NGC 6388} \\
\multicolumn{3}{c}{$\sigma(\delta$M)=0.01\msun} && \multicolumn{3}{c}{$\sigma(\delta$M)=0.01\msun} \\

\hline
        Y         &     N    &   $\delta$(M/M$_\odot$) && Y & N &   $\delta$(M/M$_\odot$) \\
\hline      
	0.33	&	90	&  0.170	  & & 0.32  &	40	 &0.185 \\  
	0.35	&	90	&  0.170	  & & 0.35  &	40	 &0.185 \\
	0.36	&	60	&  0.170	  & & 0.36	 &  40	&0.185 \\
	0.37	&	30	&  0.185	  & & 0.37  &	40&	0.195 \\
	0.38	&	20	&  0.208	  & & 0.38	 &	40&	0.204 \\
		\hline
	\end{tabular}
\end{table}

\section{Discussion}
\label{sec6}

We use new models including helium diffusion to model the evolution of the He convective regions, which reflects in the possible survival of surface turbulence in HB stars of \teff\ larger than the G--jump location. The standard location at \teff=11,500\,K corresponds to the onset of full H-ionization in the envelopes. We have shown that turbulence may survive at larger \teff\ only if models evolve from lower to larger \teff, as it only occurs at high metallicity and high helium content \citep{sweigartgross1976, caloidantona2007}.   

We have analyzed the distribution of HB stars in the two peculiar clusters NGC\,6388 and NGC\,6441 based on observations taken within the HST UV Legacy Treasury data \citep{piotto2015} an the ACS GC Survey \citep{sarajedini2007}.
This new analysis provides a further strong evidence for the presence of stars with very large initial helium content in these clusters. In particular:
\begin{enumerate}
\item We confirm the result \citep{caloidantona2007, busso2007} that a huge helium range ($\Delta$Y$\approx$0.13 in our models) is necessary to understand the extension of the HB to large \teff's, atypical for the relatively high metallicity of the stars in these clusters; a large helium spread ($\Delta$Y$\sim$0.08--0.1) is already present among the stars populating the red side of the HB. This is shown by the comparison of the tracks with the data in the clump; 
\item Y values in the range 0.34--0.38 for the tracks crossing the RR\,Lyr region provide periods consistent with  the long periods of the RR\,Lyrs in these clusters;
\item the discontinuity in the distribution of stars at \teff$\sim$7500\,K is a result of the rapid excursion of the tracks evolving from the red to the blue side of the HB, and thus a further signal of high Y at these \teff's.
\item HB tracks including helium atomic diffusion and a parametric treatment of metal levitation model the interaction between the timescales of diffusion of helium and the timescale of the \teff\ evolution of the tracks. 
\item the analysis of the blue HB side shows that we expect that the G--jump is occurring at \teff\ larger than the standard G--jump \teff, due to the peculiar shape of high--helium evolutionary tracks. 
High--Y high--Z models cross the color magnitude diagram from low to high \teff\ \citep{sweigartgross1976, caloidantona2007}, on a timescale short enough to allow the survival of the helium partial ionization regions at \teff$>$11,500\,K; this is an additional proof that the blue HB subset of stars in NGC\,6388 and NGC\,6441 have large helium abundance; 
\item for the set of computed models (Z=0.006 and [$\alpha$/Fe]=0.4), the helium contents which may justify the presence of the hotter G--jump are in the range Y=0.36--0.38;
\item the data for NGC\,6388 show also the presence of a standard  G--jump at \teff$\sim$11,500\,K \citep{brown2016}. If this is the case, some of the stars in between the two G--jumps may have achieved full helium settling. This is consistent with the simulations of stars with initial Y=0.35 - 0.36 (panel b-c of Fig.\,7) crossing the \teff\ region between the two jumps;
\item observations aimed at measuring the helium and metal content in the atmospheres of HB stars in the range 11,500 $<$ \teff $<$ 13,500\,K may confirm this model;
\item in all other galactic GCs, the metallicity is generally smaller and the HB tracks evolve towards the G--jump from hotter \teff\ for any Y value. Models show that helium is substantialy depleted and the envelope is radiatively stable in these stars, so that there is a sudden onset of convection below \teff=11,500\,K, when  hydrogen is partially recombined. The radiative levitation of metals sharply ends at this single \teff, whatever the initial Y value of these stars. This is probably the reason why the G--jump location in all cluster occurs at the same \teff \  \citep{brown2016}, in spite of the possible differences in the initial helium abundance of the second generation stars at the G--jump.
\end{enumerate}

\section*{Acknowledgments}

Support for program GO-13297 was provided by NASA through a grant from the Space Telescope Science Institute, which is operated by the Association of Universities for Research in Astronomy, Inc., under NASA contract NAS 526555. 

M.T., F.DA., S.C. and G.P. acknowledge partial support by PRIN-INAF 2014 (PI: S. Cassisi). 
G.P. also acknowledges partial support by the Università degli Studi di Padova - Progetto di Ateneo CPDA141214 ''Towards understanding complex star formation in Galactic globular clusters''.

M.D.C. acknowledges the contribution of the FP7 SPACE project ASTRODEEP (Ref. No. 312725), supported by the European Commission.

A.P.M. and A.F.M. have been supported by the Australian Research Council through Discovery Early Career Researcher Awards DE150101816 and DE160100851

\bsp	
\bibliographystyle{mnras}
\bibliography{hbtailo-revision}
\end{document}